Submitted to the Journal of the Washington Academy of Sciences on 09 March 2021        1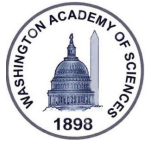

# LOW-FREQUENCY TWO-METER SKY SURVEY
# RADIAL ARTIFACTS IDENTIFIED AS BROADLINE QUASARS

ANTONIO J. PARIS
PLANETARY SCIENCES, INC.## ABSTRACT

Through the use of a High Band Antenna (HBA) system, the Low Frequency Array (LOFAR) Two-Meter Sky Survey (LoTSS) is an attempt to complete a high-resolution 120-168 MHz survey of the northern celestial sky. To date, $\sim 5 \times 10^5$ radio sources have been classified by LOFAR with most of them consisting of active galactic nuclei (AGNs). The strong AGN emissions detected by LoTSS, it is thought, are powered by supermassive black holes (SMBHs) at the center of galaxies. During an analysis of ~1,500 images of these AGNs, we identified 10 radio sources with radial spokes emitted from an unknown source. According to LOFAR, the radial spokes are artifacts due to calibration errors with no origin, and therefore they cannot be associated with an optical source. Our preliminary hypothesis for the artifacts was that they were produced by ionized jets emitted from quasi-stellar objects (QSOs). Specifically, that strong emissions from Type 1 broadline (BL) quasars were directed in the line of sight of the observer (i.e., LOFAR) and as a result produced the image artifacts. To test our hypotheses, we cross-referenced the *ra* and *dec* coordinates of the artifacts with the galactic coordinates indexed in the Sloan Digital Sky Survey (SDSS) and confirmed that the artifacts were associated with BL QSOs. Further analysis of the QSOs, moreover, demonstrated they exhibited prominent broad emission lines such as $C_{III}$ and $Mg_{II}$, which are characteristic of Type 1 BL quasars. It is our interpretation, therefore, that the radial spokes characterized as artifacts by LOFAR were produced by the emission of Type 1 BL quasars in the line of sight of the radio telescope.## INTRODUCTION

LoTSS is an ongoing high-resolution 120-168 MHz survey of the northern hemisphere designed (1) to associate or separate cataloged radio components into distinct radio sources and (2) to identify and characterize their optical counterparts.[i] Through the use of LOFAR's HBA system, LoTSS may reach a sensitivity of less than 0.1 mJy beam -1 at an angular resolution of ~2 meters across the northern hemisphere.[ii] The primary scientific objectives of the radio survey are to explore the formation and evolution of $\sim 10^7$ radio sources consisting of mostly star-forming galaxies, galaxy clusters, and SMBHs.[iii] The LOFAR source finder computer program, however, inadvertently split hundreds of thousands of the radio sources into multiple separate components. In an effort to reconstruct the full radio sources from their separated components, scientists at LOFAR created the *Radio Galaxy Zoo: LOFAR Project* and solicited the assistance of volunteer citizen scientists to classify and index the radio sources visually.[iv] As of August 2020, the survey had covered ~3,000 square degrees of the sky and cataloged $\sim 5 \times 10^5$ radio sources, and staff at LOFAR had analyzed ~80,000 of them.[v]

## ACTIVE GALACTIC NUCLEI

The majority of known radio sources in the LOFAR radio survey are AGNs.[vi] These types of active galaxies have a compact region at their center with a higher-than-normal luminosity over at least some portion of the electromagnetic spectrum when compared to an otherwise typical galaxy.[vii] The excess non-stellar emissions from these extragalactic emitters can be observed in the radio band (i.e., LOFAR), as well as through microwave, infrared, optical, ultra-violet, X-ray, and gamma ray wavebands. The non-stellar radiation emitting from AGNs is theorized to result from the accretion of matter by a SMBH at the center of its host galaxy.[viii]

In some instances, the SMBHs of AGNs produce narrow beams of high-energetic particles and eject them outward in opposite directions away from the disk (Figure 1a). These powerful jets, which emerge at nearly the speed of light, spread out up to hundreds of kiloparsecs outside the host galaxies and form distinctive radio lobe structures commonly referred to as "straddling the galaxy" (Figure 1b and 1c).[ix] The lobes are often connected by a narrower bridge of radio emission produced by the jets. If the narrow jets are visible, the optical counterpart source is generally halfway between the two lobes, and it can be cross-referenced with other spectral databases, such as SDSS (Figure 1d).[x]

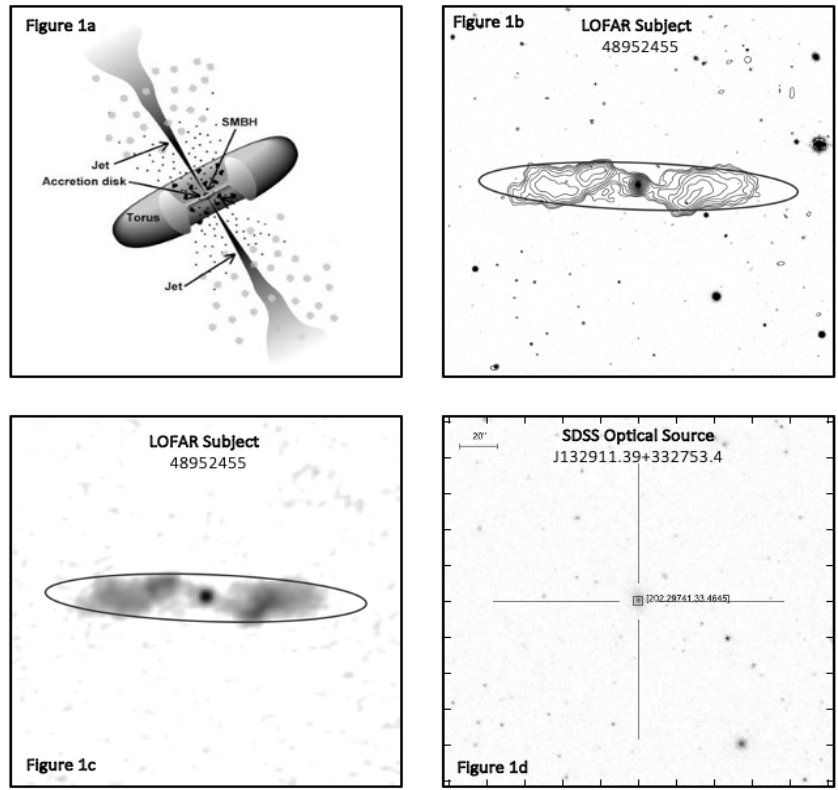

Figure 1: A NASA illustration of an AGN emitting ionized jets (1a) compared to an AGN (subject 48952455) cataloged by LOFAR (1b with artificial counters, 1c radio). The AGN's optical counterpart, identified as a starforming galaxy, is indexed in the SDSS database (1d).

## LOFAR ARTIFACTS

On 19 February 2019, LOFAR released the first dataset (DR1) collected by LoTSS, which covered 424 square degrees of the northern sky and included over ~$3 \times 10^5$ radio sources.[xi] After numerous visual sorting stages, ~2,543 of the LoTSS images were flagged with artifacts. These artifacts, referred to by LOFAR staff and several citizen scientists as "explosions," are generally near the vicinity of bright sources in the LoTSS radio image (Figure 2a and 2b). According to LOFAR, the dynamic range limitations in the imaging are responsible for the artifacts. Moreover, per the LOFAR Field Guide, bright radio sources cause radio contours to appear that have no real origin and therefore cannot be associated with an optical source. The images with the artifacts were dropped from further analysis by staff at LOFAR and were not included in the final LoTSS-DR1 catalogue. Furthermore, no clarification regarding the direct source of the artifacts other than an unknown bright source was provided.

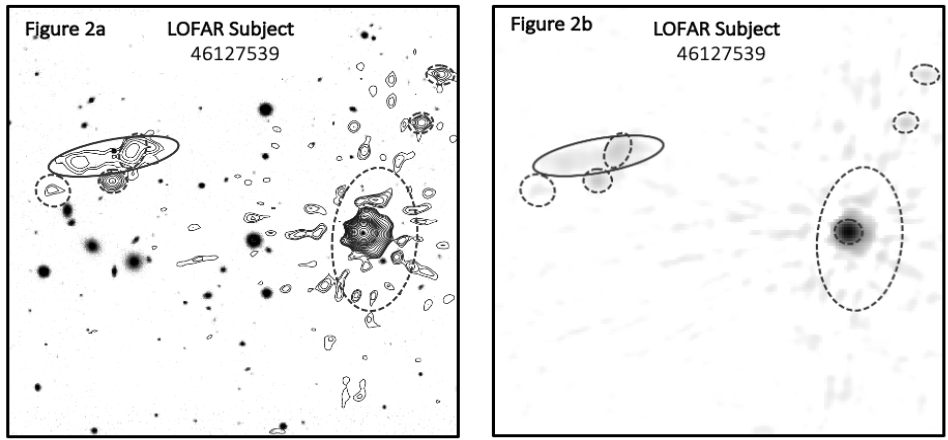

Figure 2: An example of a radio image with artifacts observed by the LOFAR survey (2a with artificial counters, 2b radio). The artifacts, according to LOFAR, appear as radial "explosions" emitted from an unknown bright source located at ra: 352.284369, dec: 18.576822.

## QUASI STELLAR OBJECTS

Our preliminary hypothesis of the source of the LOFAR artifacts was quasars. Specifically, we hypothesized that the ionized jets emitted from a radio-load quasar were in the line of sight of the LOFAR radio telescope, and, as a result, produced the artifacts. Quasars, also referred to as quasi-stellar objects, are the most powerful subclass of AGNs. They have strong radio emissions and optically have a starlike visual appearance.[xii] The radiation emitted by SMBHs at the center of quasars is among the most luminous, powerful, and energetic phenomena known in the universe.[xiii]

Quasars are further defined based on their observed characteristics, such as how the ionized jets and gases emitted from the disk are oriented to the observer (Figure 3). Radio-load quasars have the most powerful jets and are strong sources of radio-wavelength emission. In contrast, radio-quiet quasars have weaker yet still powerful jets, with relatively weaker radio emissions from the SMBHs.[xiv] Blazars, which are rare, are radio-loud quasars with relativistic jets directed in the line of sight of an observer.[xv]

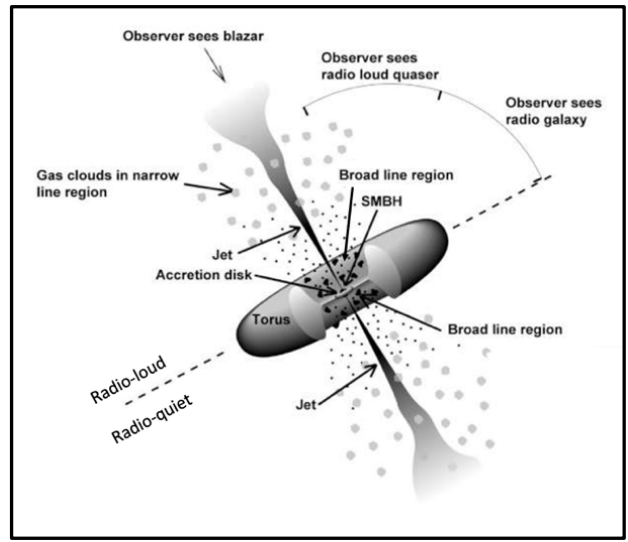

Figure 3: An illustration demonstrating how the orientation of emissions from AGNs are defined (source: NASA).

BL quasars are a subset of quasars whose optical spectra emission lines are believed to come from high-velocity gas in the broadline region (BLR), ~0.1-1 pc from the central engine of the SMBH (Figure 3).[xvi] BL quasars are further categorized as Type 1 objects because they display broad lines in their unpolarized spectra, have a low inclination angle with respect to the accretion disk axis, and consequently enable an unobscured view of the central engine.[xvii] The optical spectra of BL quasars, moreover, exhibit prominent broad emission lines but rather weaker narrow lines. Prominent broad emission lines in BL quasars include $C_{III}$ and $Mg_{II}$. Moreover, $L_{ya}$ emission lines arise from the electronic transition in atomic hydrogen from the state n = 2 to n = 1.[xviii] This line, which occurs at a wavelength of 121.6 nm, typically indicates very strong Type 1 BL quasars.

## CROSS-MATCHING LOFAR ARTIFACTS WITH
## THE SLOAN DIGITAL SKY SURVEY (SDSS)

Our assumption regarding the source of the LOFAR artifacts is they are associated with ionized gas emitted from the BLR of quasars. To corroborate our hypotheses, we cross-matched the *ra* and *dec* coordinates of 10 LOFAR radio sources (with artifacts) with SDSS—a multi-spectral imaging survey that uses a 2.5-m wide-angle optical telescope.[xix] Cross-matching radio signatures with SDSS is a well-established procedure in astronomy. For many radio sources, such as quasars, emissions are relatively compact and are analogous with their optical counterparts, allowing unpretentious cross-matching. The optical counterpart for LOFAR radio subject 44118344 (Figure 4a), for illustration, was indexed in SDSS (Figure 4b) and, as we suspected, was classified as a Type 1 BL quasar. The broad emission lines $Mg_{II}$ in optical and $C_{III}$ in the near infrared, which are characteristic of Type 1 BL quasars, are also visible in the SSDS spectra (Figure 4c), supporting our hypothesis that the emission from 44118344 came from a BL quasar. Likewise, when we cross-matched the remaining nine LOFAR subjects (with artifacts) in SDSS, we confirmed they too were indexed as Type 1 BL quasars (Table 1 and Appendix 1) with accompanying spectra (Table 2) characteristic of these objects.

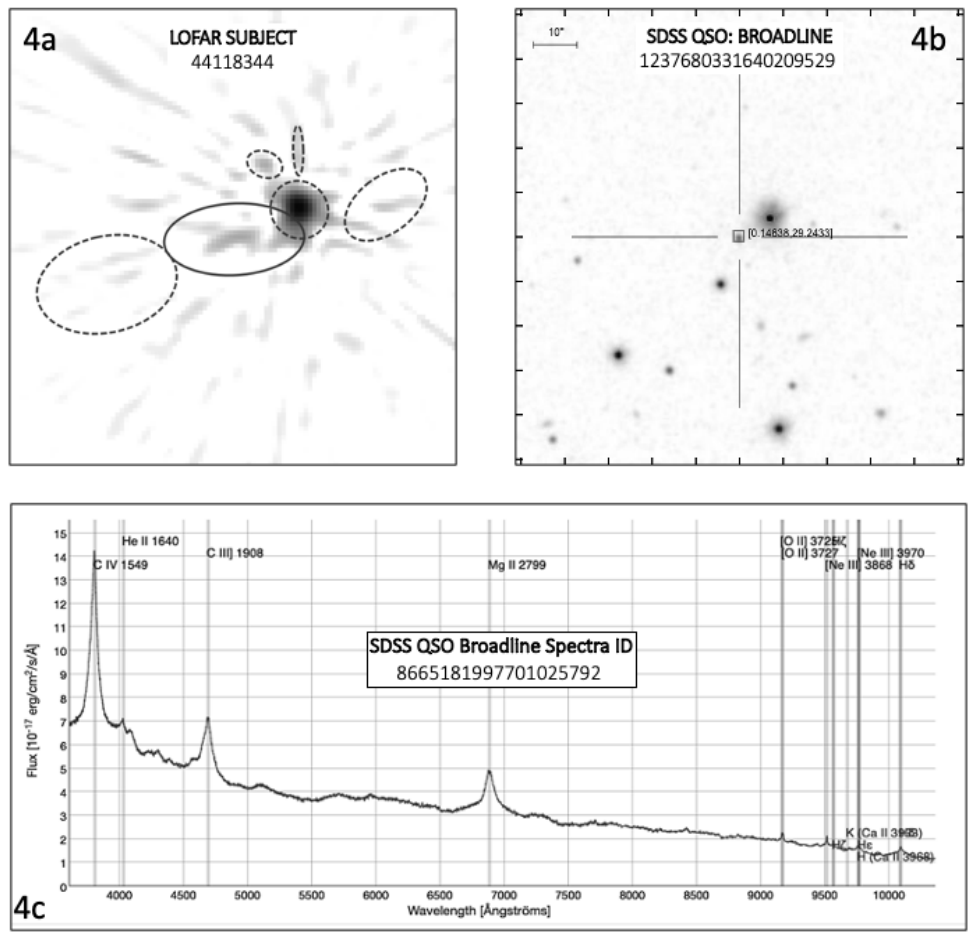

Figure 4: A crossmatch of LOFAR subject 44118344 (4a) with a BL quasar indexed in SDSS (4b). The spectra (4c) fit a BL quasar.

| LOFAR Artifact | SDSS Object ID | SSDS Sprectral ID | Quasar Type | Prominent Broad Emission Lines Wavelength [Ångströms] | | | Redshift $z$ |
|---|---|---|---|---|---|---|---|
| | | | | Ly $a$ | C III 1980 | MG II 2799 | |
| 44118344 | 1237680331640209529 | 8665181997701025792 | BLQ | | 4690.294 | 6883.353 | 1.459 |
| 44120630 | 1237673072594911468 | 7352158936949346304 | BLQ | 3922.835 | 6156.019 | 9019.886 | 2.228 |
| 44209850 | 1237662474227810333 | 1594412579910871040 | BLQ | | 3821.222 | 5447.536 | 0.941 |
| 47773327 | 1237678601851502861 | 5995206854356742144 | BLQ | | 3932.789 | 5768.994 | 1.059 |
| 44128911 | 1237680301029065381 | 8537969040429436928 | BLQ | 3610.772 | 5678.862 | 8311.892 | 1.969 |
| 50911574 | 1237680332723126683 | 7053769068508368896 | BLQ | | 3682.136 | 5395.108 | 0.928 |
| 48239351 | 1237660962404106339 | 8468973069483618304 | BLQ | | 3815.926 | 5591.135 | 0.998 |
| 48192498 | 1237664668433318086 | 6926359617319186432 | BLQ | | 3729.065 | 5451.299 | 0.946 |
| 44128422 | 1237680329484272602 | 8609971395701788672 | BLQ | | 5476.463 | 8011.247 | 1.865 |
| 46127539 | 1237680245200585117 | 8561389187672330240 | BLQ | | 3709.368 | 5470.159 | 0.952 |

Table 2: 10 LOFAR radio artifacts with spectra characteristic of BL quasars (source: SDSS).

| LOFAR Artifact ID | LOFAR Ra | LOFAR Dec | SDSS Object ID | SDSS Subject ID | Plate | Quasar Type | Galactic Coordinates $l$ | Galactic Coordinates $b$ | Redshift $z$ |
|---|---|---|---|---|---|---|---|---|---|
| 44118344 | 0.153533 | 29.241163 | 1237680331640209529 | J000035.13+291435.8 | 7696 | BLQ | 109.797043 | -32.3280304 | 1.459 |
| 44120630 | 8.695916 | 33.488398 | 1237673072594911468 | J003445.88+333018.5 | 6530 | BLQ | 118.949151 | -29.2375322 | 2.228 |
| 44209850 | 236.658701 | 36.740359 | 1237662474227810333 | J154638.31+364420.2 | 1416 | BLQ | 58.772973 | 51.9544549 | 0.941 |
| 47773327 | 22.497927 | 17.102729 | 1237678601851502861 | J013003.32+170619.5 | 11059 | BLQ | 135.983923 | -44.7853478 | 1.059 |
| 44128911 | 336.613764 | 23.346522 | 1237680301029065381 | J222619.60+232011.3 | 7583 | BLQ | 84.718632 | -28.5644715 | 1.969 |
| 50911574 | 24.196933 | 28.678306 | 1237680332723126683 | J013639.88+284012.5 | 6265 | BLQ | 134.789731 | -33.1580088 | 0.928 |
| 48239351 | 141.468759 | 36.213063 | 1237660962404106339 | J092551.85+361235.6 | 8862 | BLQ | 187.770071 | 45.8434456 | 0.998 |
| 48192498 | 150.318035 | 34.429497 | 1237664668433318086 | J100111.95+342450.4 | 10232 | BLQ | 190.890222 | 52.9883760 | 0.946 |
| 44128422 | 334.433215 | 29.643024 | 1237680329484272602 | J221742.67+294027.5 | 7647 | BLQ | 87.209568 | -22.3334179 | 1.865 |
| 46127539 | 352.284369 | 18.576822 | 1237680245200585117 | J232903.37+183430.7 | 7604 | BLQ | 97.078260 | -40.1246700 | 0.952 |

Table 1: A sampling of 10 LOFAR radio source artifacts cross-matched and confirmed as BL quasars indexed in SDSS.

## CONCLUSIONS

LoTSS has made considerable gains in achieving its goal of cataloging and cross-matching thousands of AGNs in the northern hemisphere. In some instances, however, the limitations of the two-meter telescope resulted in artifacts that were not thoroughly identified by LOFAR. To identify the source of the artifacts, we examined 10 LOFAR radio sources that exhibited artifacts and cross-matched them with the multi-spectral imaging survey SDSS. As we initially suspected, we identified the source of the artifacts as radio-loud quasars. Moreover, scrutiny of the data indexed in SDSS showed that the LOFAR telescope was in the line of sight of the BLR of the 10 quasars. We make the latter assumption from an analysis of the accompanying spectra in SDSS, which exhibited prominent broad emission lines $C_{III}$ and $Mg_{II}$, which are characteristic of Type 1 BL quasars. It is our interpretation, accordingly, that the radial spokes categorized as artifacts by LOFAR were produced by the emission of QSOs.

## NOTES

The Radio Galaxy Zoo: LOFAR Project is part of Radio Galaxy Zoo (RGZ)—an internet crowdsourced citizen science project attempting to locate SBMHs in distant galaxies. The programs are hosted by the web portal Zooniverse. Using many classifications provided by citizen scientists, as well as professional scientists, it hopes to build a more complete picture of black holes at various stages and their origins.[xx]

## BIOGRAPHY

Antonio Paris, the Principal Investigator (PI) for this study, is the Chief Scientist at Planetary Sciences, Inc., an Assistant Professor of Astronomy and Astrophysics at St. Petersburg College, FL, and a graduate of the NASA Mars Education Program at the Mars Space Flight Center, Arizona State University. He is the author of *Mars: Your Personal 3D Journey to the Red Planet*. His latest peer-reviewed publication is *Prospective Lava Tubes at Hellas Planitia*—an investigation into leveraging lava tubes on Mars to provide crewed missions with protection from cosmic radiation. Prof. Paris is a professional member of the Washington Academy of Sciences, the American Astronomical Society, and the Explorer's Club.

## CONTRIBUTIONS

Dennis Farr, who currently serves as President at the Society of Amateur Radio Astronomers, assisted with cross-matching SDSS data during investigation.

# APPENDIX 1

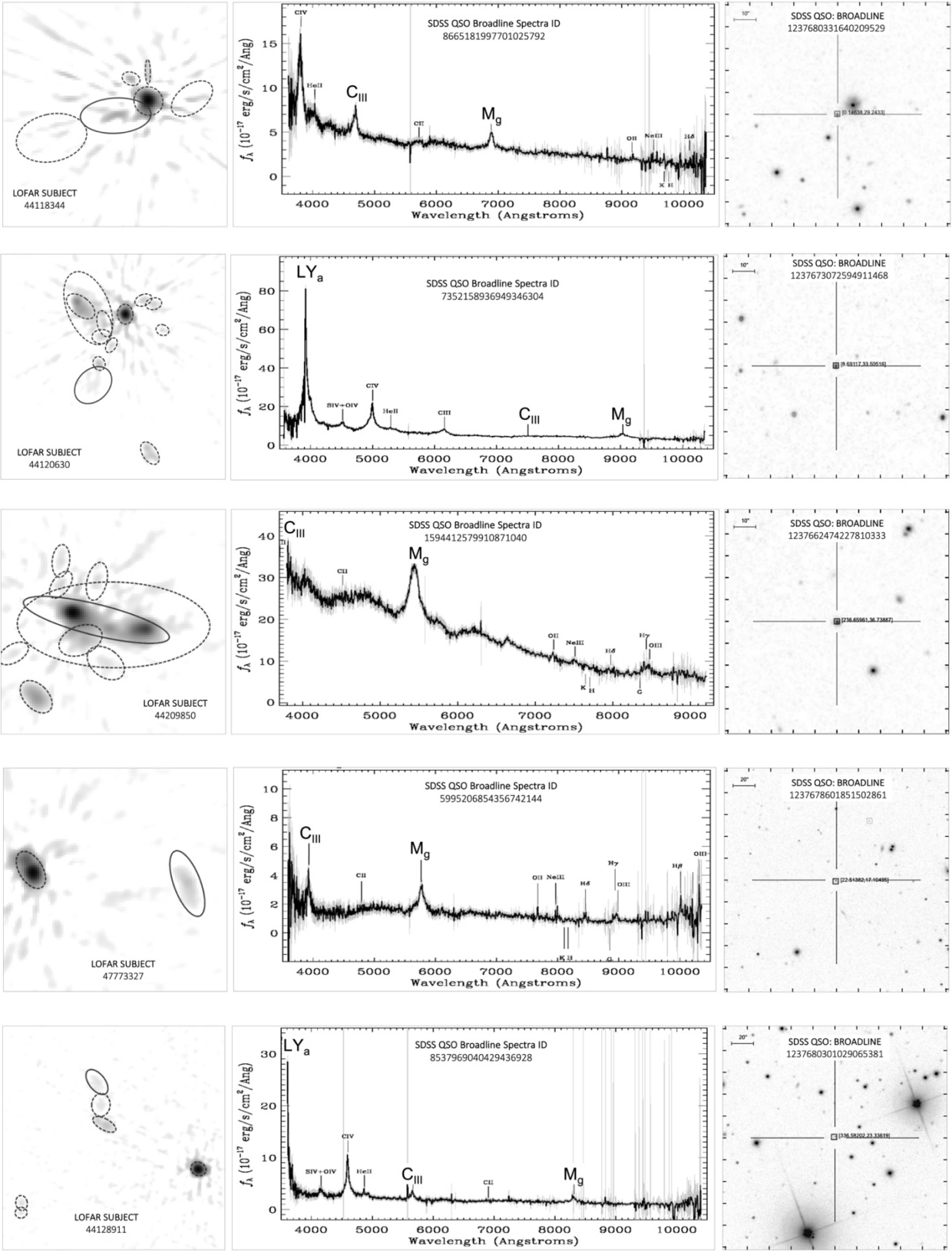

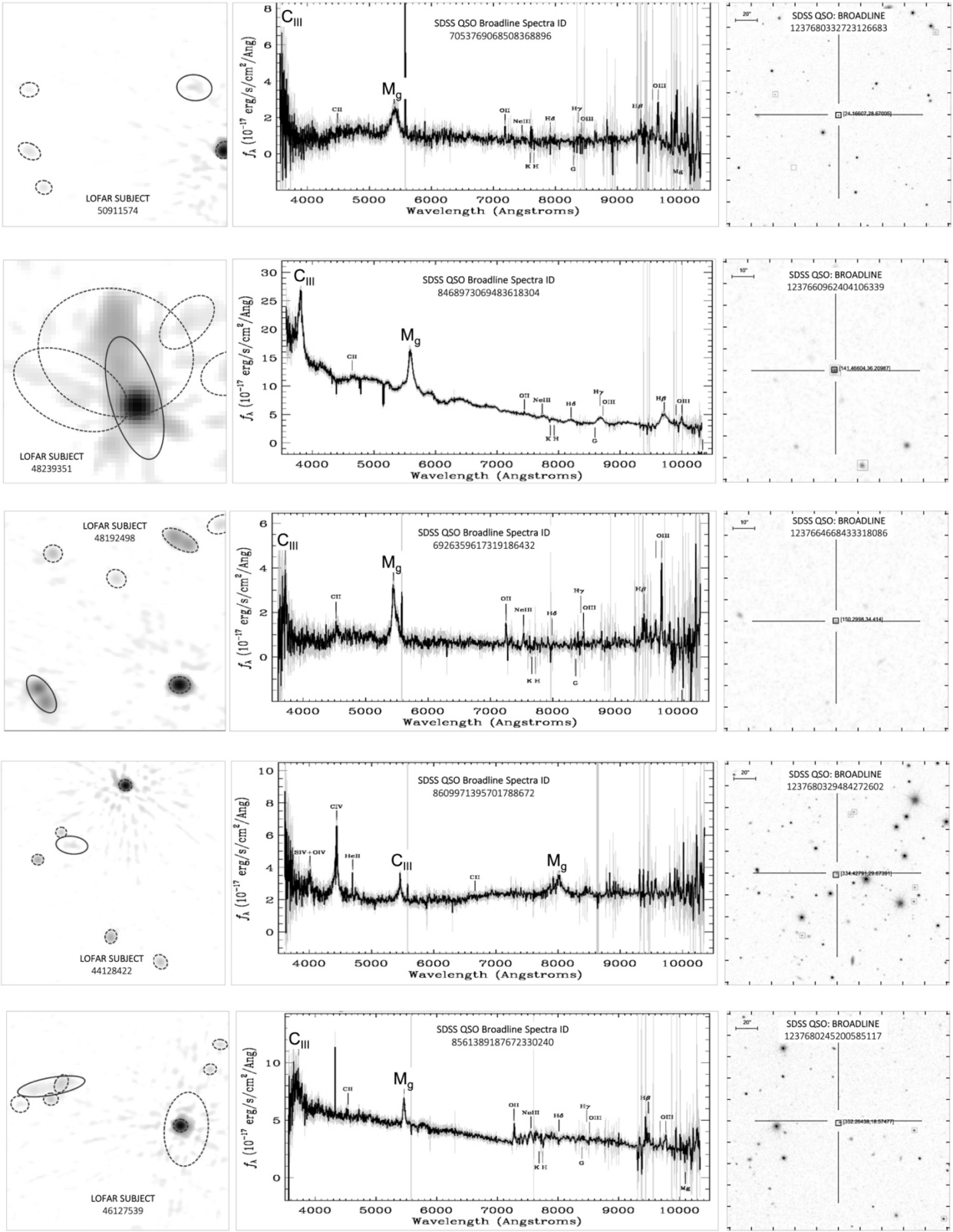